\documentclass[10pt, final, journal, twoside]{IEEEtran}

\usepackage{multicol}

\usepackage[margin=0.5in]{geometry}

\usepackage{upgreek} 
\usepackage{dblfloatfix} 
\usepackage{caption}
\usepackage{cite}
\usepackage{mathtools}
\usepackage{epsf}
\usepackage{commath}
\usepackage{amsmath}
\usepackage{amssymb}
\usepackage{amsfonts}
\usepackage{graphicx}
\usepackage{mwe,tikz}
\usepackage[percent]{overpic}
\usepackage{psfrag}
\usepackage{array,multirow,graphicx}
\usepackage{newunicodechar}
\usepackage{subfigure}
\DeclareMathOperator{\sinc}{sinc}
\newcommand{\re}{r_{\mathrm{e}}}
\newcommand{\phimax}{\phi_{\mathrm{max}}}
\newcommand{\XMIN}{X_{\mathrm{min}}}

\newcommand{\Gammamin}{\Upupsilon_{\mathrm{min}}}

\newcommand{\alphamax}{\alpha_{\mathrm{max}}}

\newcommand{\alphamin}{\alpha_{\mathrm{min}}}

\newcommand{\fd}{\chi_{\mathrm{t}}}

\newcommand{\GR}{G_\mathrm{R}}

\newcommand{\GMAX}{G_\mathrm{max}}

\newcommand{\HS}{H_\mathrm{S}}

\newcommand{\PTX}{\mathrm{P}_{\mathrm{tx}}}

\usepackage{scalerel}

\usepackage{tikz}
\usetikzlibrary{svg.path}
\newcounter{MYtempeqncnt}

\definecolor{orcidlogocol}{HTML}{A6CE39}
\tikzset{
  orcidlogo/.pic={
    \fill[orcidlogocol] svg{M256,128c0,70.7-57.3,128-128,128C57.3,256,0,198.7,0,128C0,57.3,57.3,0,128,0C198.7,0,256,57.3,256,128z};
    \fill[white] svg{M86.3,186.2H70.9V79.1h15.4v48.4V186.2z}
                 svg{M108.9,79.1h41.6c39.6,0,57,28.3,57,53.6c0,27.5-21.5,53.6-56.8,53.6h-41.8V79.1z M124.3,172.4h24.5c34.9,0,42.9-26.5,42.9-39.7c0-21.5-13.7-39.7-43.7-39.7h-23.7V172.4z}
                 svg{M88.7,56.8c0,5.5-4.5,10.1-10.1,10.1c-5.6,0-10.1-4.6-10.1-10.1c0-5.6,4.5-10.1,10.1-10.1C84.2,46.7,88.7,51.3,88.7,56.8z};
  }
}

\newcommand\orcidicon[1]{\href{https://orcid.org/#1}{\mbox{\scalerel*{
\begin{tikzpicture}[yscale=-1,transform shape]
\pic{orcidlogo};
\end{tikzpicture}
}{|}}}}

\usepackage[hidelinks]{hyperref} 

\begin{document}

\title{Performance Analysis of LEO-Terrestrial Systems in Presence of Doppler Effect}
\author{Islam~M.~Tanash$^{\orcidicon{0000-0002-9824-6951}}$, Nuria González-Prelcic$^{\orcidicon{0000-0002-0828-8454}},$~\IEEEmembership{Fellow,~IEEE}, and Risto Wichman$^{\orcidicon{0000-0002-5261-5037}},$~\IEEEmembership{Senior Member,~IEEE}
\vspace{-1.1cm}%
\thanks{Islam M. Tanash is with the Department of Electrical Engineering, Prince Mohammad Bin Fahd University, Al Khobar 31952, Saudi Arabia (e-mail: itanash@pmu.edu.sa).}

\thanks{Nuria Gonz\'alez-Prelcic is with the Department of Electrical and Computer
Engineering, University of California San Diego, La Jolla, CA 92093 USA
(e-mail: ngprelcic@ucsd.edu).}

\thanks{Risto Wichman is with the Department of Information and Communications Engineering, Aalto University, 00076 Espoo, Finland (e-mail: risto.wichman@aalto.ﬁ).}
\thanks{The work was supported by Research Council of Finland Grant 339446}
}%

\maketitle
\vspace{-.9cm}

\begin{abstract}
In this paper, we present a novel stochastic geometry-based approach to analyze the effect of residual Doppler shift on {orthogonal frequency-division multiple access (OFDMA)} systems in {low earth orbit (LEO) satellite-terrestrial networks. Focusing on multiuser systems employing common Doppler compensation, we analytically formulate the coverage probability by explicitly capturing the loss of OFDMA subcarrier orthogonality caused by geometry-induced residual Doppler through inter-carrier interference.} The analysis accounts for the spatial distribution of ground terminals within the serving satellite’s cell and is validated through extensive Monte--Carlo simulations for both S-band and Ka-band settings. The results demonstrate the high accuracy of both the Doppler shift approximation and the derived coverage probability expression, while also highlighting the significant impact of residual Doppler shift, even after compensation, emphasizing the necessity of considering this effect in the design of future satellite networks.

\end{abstract}
\vspace{-.6 cm}

\section{Introduction}
\label{sec:introduction}
Recent research has focused on low Earth orbit (LEO) satellite communication systems due to their advantages over medium Earth orbit (MEO) and geostationary orbit (GEO) systems, particularly because of their lower power consumption and reduced latency. LEO satellite communications are particularly well-suited for providing connectivity in underserved regions, such as rural and remote areas lacking infrastructure~\cite{general_leo_survey}, or complementing terrestrial communication in urban areas~\cite{Tanash_2024}. The 3GPP's proposal of incorporating non-terrestrial networks into the 5G framework underscores the significance of LEO satellites~\cite{TR38.821}. Such LEO systems offer a solution to the substantial delays found in the communication links of GEO satellites. However, they also present challenges, with high mobility being a major concern that leads to increased Doppler shifts, potentially affecting their performance.

A common solution to the prominent Doppler shift problem is to use frequency compensation techniques at a reference point within the satellite's coverage area, by which the Doppler shift is compensated similarly for all users, thereby partially mitigating its effect~\cite{compen4,compensate,compens3}. Due to the wide geographical spread of users within the coverage area, residual Doppler shift, often referred to as differential Doppler shift, persists among users. This can cause intercarrier interference degrading 
the performance of multicarrier systems, such as orthogonal frequency-division multiple access (OFDMA) systems. For reliable and precise satellite communications, it is necessary to predict the Doppler shift within the satellite's coverage area and study the effect of the residual Doppler shift on the performance of the LEO satellite-terrestrial networks.
Several works in the literature have analytically derived the Doppler shift experienced by a ground user under deterministic conditions{~\cite{new_ref1,new_ref2,ali}}, while others have statistically characterized its behavior by modeling the randomness in network geometry, aiming to extract insightful design guidelines~\cite{akram_doppler,dop_iot,main,doppler_tanash}.


{To the best of the authors’ knowledge, the existing literature lacks system-level analyses that incorporate geometry-induced residual Doppler after common compensation into fundamental performance metrics such as coverage probability for LEO satellite–terrestrial systems. Prior works either characterize Doppler statistics in isolation or analyze link-level interference effects, but do not capture how residual Doppler differences across users sharing a common OFDMA waveform lead to loss of subcarrier orthogonality and propagate into system-level performance. 
Motivated by this gap, this letter studies the effect of residual Doppler on the performance of multiuser LEO satellite–terrestrial systems operating with OFDMA. We first derive an accurate and analytically tractable expression for the Doppler shift at an arbitrary ground location within the satellite footprint during a LEO pass, based on a flat-Earth approximation. We then statistically characterize the geometry-induced residual Doppler after common compensation across users. By reinterpreting residual Doppler as a user-dependent frequency mismatch on a shared OFDM grid, we characterize the resulting inter-carrier interference using a suitable OFDM model. Finally, we embed this impairment into the signal-to-interference-plus-noise ratio and derive a closed-form expression for the coverage probability that explicitly accounts for residual Doppler effects.}



\vspace{-.2cm}
\section{System Model}

\label{sec:system_model}
 \begin{figure}[t]
    \centering
    \includegraphics[trim=6.5cm 1.5cm 0.5cm 0.80cm, clip=true, width=0.46\textwidth ]{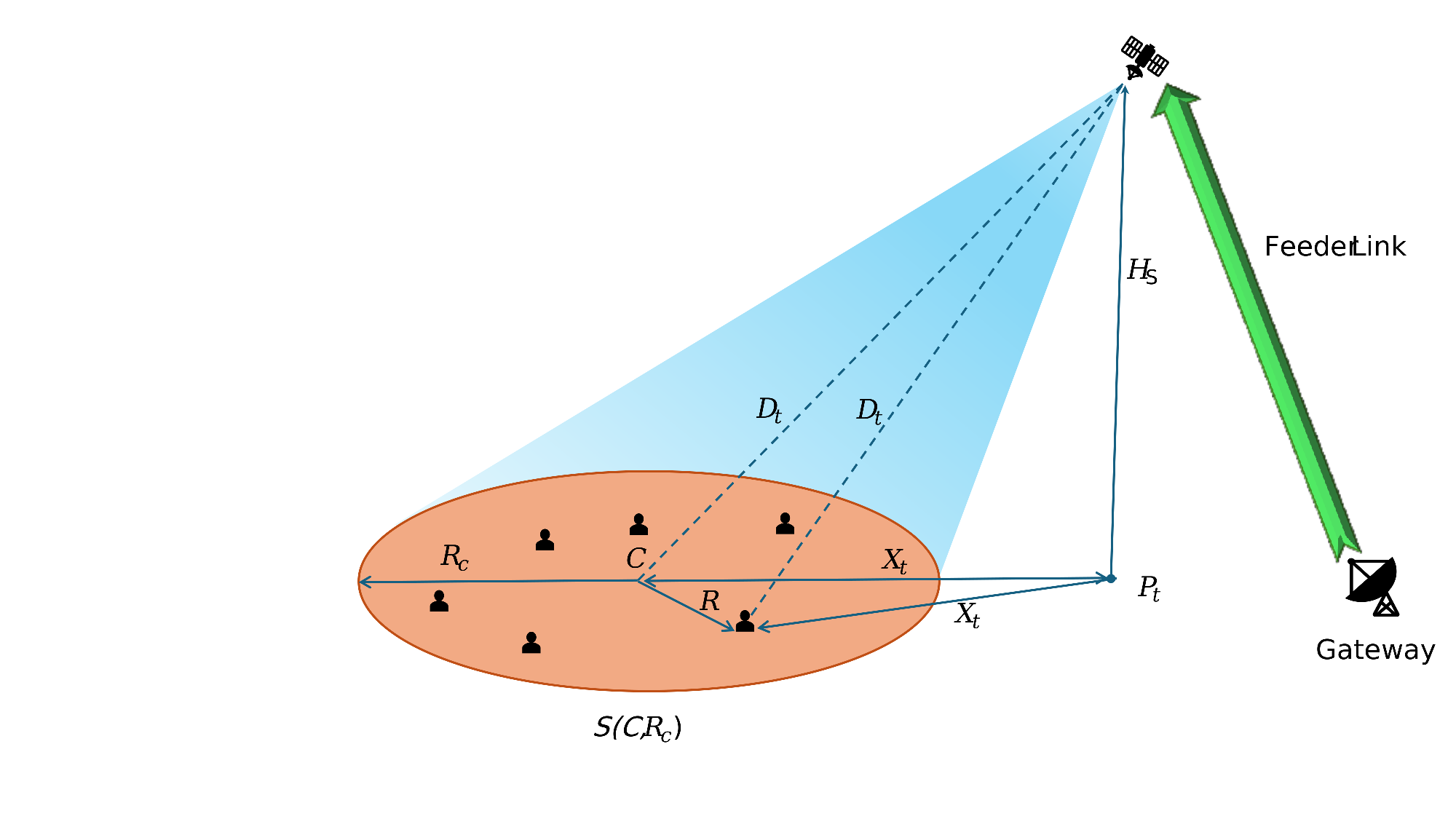}
\caption{Satellite-terrestrial communication system with randomly distributed ground terminals. The parameters shown in the figure and their descriptions are summarized in Table~\ref{table:notations}.}
    \label{fig:system_model}
    \vspace{-.5 cm}
\end{figure}

\begin{table*}[h]
\centering
\caption{Table of Notations and Descriptions}
\label{table:notations}
\begin{tabular}{c|c}
\hline Notation & Description \\
\hline \hline $C$; $R_c$; $P_t$ & Center of the circular beam; Radius of the corresponding circular cell; Satellite's Projection into Earth's surface at time $t$ \\
\hline $R$; $\HS$; $\re$ & Distance from
C to the ground terminal; Altitude of the serving satellite; Radius of Earth\\
\hline$X_t$; $D_t=\sqrt{X_t^2+\HS^2}$ & Horizontal distance between the ground terminal and $P_t$ at time $t$; Corresponding slant distance to the satellite  \\
\hline $\hat{X}_t$; $\hat{D}_t=\sqrt{\hat{X}_t^2+\HS^2}$& Horizontal distance between the cell's center $C$ and $P_t$; Corresponding slant distance from $C$ to the satellite\\
\hline $\XMIN$ & Minimum possible horizontal distance from the ground terminal to the nearest point on the serving orbit
\\
\hline$\hat{X}_{\mathrm{min}}$ & Minimum possible horizontal distance from $C$ to the nearest point on the serving orbit\\
\hline$\alphamax$ & Maximum possible elevation angle which corresponds to $\XMIN$ through $\alphamax=\arctan\left(\frac{\HS}{\XMIN}\right)$  \\
\hline  $\fd$; $\delta_t$; $\Delta \delta_t$ & Doppler shift at the ground terminal; Common Doppler shift among all ground terminals; Residual Doppler shift 
\\
\hline
\end{tabular}
\vspace{-.3 cm}
\end{table*}

As shown in Fig. \ref{fig:system_model}, we consider herein a LEO satellite network, in which a satellite at altitude $\HS$ has an Earth-fixed single beam. {Although the analysis focuses on a single spot beam for clarity, the derived framework and coverage probability characterize the performance of an individual beam and can be directly applied to each beam independently in a multi-beam satellite architecture.} The circular spot beam, when directed onto the surface of Earth, generates a circular cell of radius $R_c$, using the ﬂat-Earth approximation. This cell represents the footprint of the spot beam on the Earth's surface and is defined as $S(C,R_c)$, where $C$ is the cell's center and $A_c=\pi\,R_c^2$ is its area. We assume that ground terminals\footnote{Ground terminals may be either stationary or slowly moving. In particular, the Doppler shift is determined by the relative velocity between the satellite and the ground terminal. Since the satellite's speed is significantly higher than that of the ground terminal, the satellite's movement is the primary contributor to the Doppler shift.}, equipped with either omnidirectional antennas (e.g., handheld devices) or directional antennas (e.g., VSAT terminals) capable of tracking the satellite’s movement, are uniformly distributed at random within the satellite's cell at time instant $t$. 
The half-power beamwidth (HPBW) of the satellite's antenna, denoted by $\psi_{3\mathrm{dB}}$, can be used to approximate the radius of the circular cell $S(C, R_c)$ using basic geometry, at the instant when the satellite is positioned directly above the cell's center, as \begin{align} R_c = \HS\, \tan\left(\frac{\psi_{3\mathrm{dB}}}{2}\right). \end{align}
Although the actual footprint becomes elliptical when the satellite is inclined or located away from the cell's center, we assume a circular cell for analytical tractability.

Doppler compensation is implemented to mitigate the effect of Doppler shift caused by satellite movement in relation to the ground terminals. However, LEO satellites do not precompensate the Doppler shift for each ground terminal since those within the same beam experience different Doppler shifts due to their different locations and elevation angles to the traversing serving satellite, making simultaneous precompensation impractical. Instead, the Doppler shift is compensated at a reference point within the cell to offset the common component of the Doppler shift experienced by all its terminals. Therefore, the residual Doppler shift at each ground terminal varies with its relative position to the reference point whose Doppler shift after compensation becomes zero. The reference point is usually chosen as the center $C$ of the cell. The ground gateway is responsible for calculating the common time-varying Doppler shift relative to the reference point and exchanging this information with the satellite via a feeder link~\cite{feeder_link}. We summarize the notation followed in this paper in Table~\ref{table:notations}.

\vspace{-.3cm}

\subsection{Doppler Characterization}
The Doppler shift magnitude expression derived in~\cite[Eq.3]{doppler_tanash} is formulated in terms of two central angles, specifically, $\Upupsilon_t$, the angle between the satellite and the ground terminal at time $t$, and $\Gammamin$, the minimum possible central angle corresponding to the satellite’s closest approach to the ground terminal. These angles relate to their corresponding arc lengths via $\Upupsilon_t = \frac{X_t}{\re}$ and $\Gammamin = \frac{\XMIN}{\re}$, where $\re$ denotes the Earth's radius. Both $\Upupsilon_t$ and $\Gammamin$ are generally bounded by $\phimax = \arccos\left(\frac{\re}{\re + \HS}\right)$ which represents the maximum central angle at which a satellite can serve the user, corresponding to the satellite’s position at the user's horizon. Within this angular range, the curvature of the Earth has relatively a negligible effect, and thus, the arc lengths $X_t$ and $\XMIN$ can be accurately approximated as flat horizontal distances under the flat-Earth assumption. Consequently, the Doppler shift magnitude at the ground terminal at time $t$, denoted by $\fd$, can be computed as
\begin{align}
\label{eq:doppler_s_approx}
{\fd}(X_t,\XMIN) &=\rho\, \,\sqrt{\frac{X_t^2-{\XMIN^2}}{\big(\frac{\re\,\HS}{\re+\HS}\big)^2+\big(\frac{\re}{\re+\HS}\big)X_t^2}},
\end{align} 
where $\rho=\frac{f_o\,\re \,\omega_f}{c}$ for which $f_o$ is the carrier frequency, $c$ is the speed of light and $\omega_F=\omega_s-\omega_E\,\cos(i)$, with $\omega_s$ and $\omega_E$ respectively representing the satellite's and Earth's angular velocities in the Earth-centered inertial (ECI) coordinate system, while $i$ represents the constellation's inclination~\cite{ali}. The velocities $\omega_s$ and $\omega_E$ are typically constant in the ECI frame with $\omega_s=\sqrt{\frac{\mu}{(\re+h)^3}}$, for which $\mu$ is the standard gravitational parameter of the Earth.
The cumulative distribution function (CDF) of the horizontal distance $X_t$ between a random ground terminal within $S(C,R_c)$ and the projection of the satellite at instant $t$ is defined in~\cite[Eq. 11]{main} and restated in (\ref{eq:CDF_Z}) at the beginning of the next page\stepcounter{equation}.

\begin{figure*}[t]
\setcounter{MYtempeqncnt}{\value{equation}}
\setcounter{equation}{2}
\begin{equation}
\resizebox{.85\hsize}{!}{$\begin{aligned}
\label{eq:CDF_Z}
F_{X_t}(x_t, R_c \mid \hat{X}_t)&=\begin{cases}\frac{x_t^2}{R_c^2}, 0 \leq x_t \leq R_c-\hat{X}_t \\ \frac{x_t^2}{\pi R_c^2}\left(\theta^*(x_t)-\frac{1}{2} \sin \left(2 \theta^*(x_t)\right)\right)+\frac{1}{\pi}\left(\phi^*(x_t)-\frac{1}{2} \sin \left(2 \phi^*(x_t)\right)\right),|R_c-\hat{X}_t|<x_t \leq R_c+\hat{X}_t
\end{cases}
\end{aligned}$}
\end{equation}
\setcounter{equation}{\value{MYtempeqncnt}}
\hrulefill\\
$^*$note: $\theta^*(x_t) =\cos ^{-1}\left(\frac{x_t^2+\hat{X}_t^2-R_c^2}{2 \hat{X}_t x_t}\right)$ and $ \phi^*(x_t) =\cos ^{-1}\left(\frac{R_c^2+\hat{X}_t^2-x_t^2}{2 \hat{X}_t R_c}\right).$
\vspace{-.3cm}
\end{figure*}

\begin{figure*}[!b]
\setcounter{MYtempeqncnt}{\value{equation}}
\setcounter{equation}{8}
\begin{equation}
\resizebox{.55\hsize}{!}{$\begin{aligned}\label{eq:cov_prob}
&P_{\mathrm{cov}}(\tau)=
\mathbb{P}\left(\mathrm{SINR}>\tau\right)\\
&= \mathbb{P}\bigg(\frac{{A }\sinc^2(|\Delta \delta_t(X_t,\XMIN)| T)\, ({X_t^2+\HS^2})^{-1}}{{A} \big(1-\sinc^2(|\Delta \delta_t(X_t,\XMIN)| T)\big)\,({X_t^2+\HS^2})^{-1}+N_o}>\tau\bigg)\\
&=\mathbb{P}\bigg(\sinc^2(|\Delta \delta_t(X_t,\XMIN)| T)>\frac{\tau+\big(\tau\,N_o\,({X_t^2+\HS^2})\big)/A}{\tau+1}\bigg)\\
&\stackrel{\text{(a)}}{=}\mathbb{P}\Bigg(\Bigg{|}{\rho\, \,\sqrt{\frac{X_t^2-{\hat{X}_{\mathrm{min}}^2}}{\big(r_o\,H_S\big)^2+r_o\,X_t^2}}-\delta_t}\Bigg{|}
 <\frac{1}{T}\text{sinc}^{-1}\Bigg(\sqrt{\frac{\tau+\big(\tau\,N_o\,({X_t^2+\HS^2})\big)/A}{\tau+1}}\Bigg)\Bigg) \\
  &=\mathbb{P}\Bigg(-\frac{1}{T}\text{sinc}^{-1}\big(\Phi\big)<{\rho\, \,\sqrt{\frac{X_t^2-{\hat{X}_{\mathrm{min}}^2}}{\big(r_o\,H_S\big)^2+r_o\,X_t^2}}-\delta_t}
 <\frac{1}{T}\text{sinc}^{-1}\big(\Phi\big) \Bigg)\\
 &=\mathbb{P}\Bigg(\sqrt{
\frac{\psi_1\,(r_o\,H_S)^2+\hat{X}_{\mathrm{min}}^2}{1+\Psi\,r_o}}< X_t < \sqrt{
\frac{\psi_2\,(r_o\,H_S)^2+\hat{X}_{\mathrm{min}}^2}{1-\Psi\,r_o}}\Bigg)\\
&=F_{X_t}\bigg(\sqrt{
\frac{\psi_2\,(r_o\,H_S)^2+\hat{X}_{\mathrm{min}}^2}{1-\Psi\,r_o}}, R_c \mid \hat{X}_t\bigg)-F_{X_t}\bigg(\sqrt{
\frac{\psi_1\,(r_o\,H_S)^2+\hat{X}_{\mathrm{min}}^2}{1+\Psi\,r_o}}, R_c \mid \hat{X}_t\bigg)
\end{aligned}$}
\end{equation}
\setcounter{equation}{\value{MYtempeqncnt}}
\hrulefill\\
\small$^*$note: $ \delta_t=\fd(\hat{X}_t,\hat{X}_{\mathrm{min}})$, $r_o=\frac{\re}{\re+\HS}$, $\Phi=\sqrt{{\frac{\tau+\big(\tau\,N_o\,({\HS^2})\big)/A}{\tau+1}}}$, $\Psi_1=\bigg(-\frac{\text{sinc}^{-1}(\Phi)}{T\,\rho}+\frac{\delta_t}{\rho}\bigg)^2$, and $\Psi_2=\bigg(\frac{\text{sinc}^{-1}(\Phi)}{T\,\rho}+\frac{\delta_t}{\rho}\bigg)^2$.
\end{figure*}

According to~\cite[Remark 1]{doppler_tanash}, different UEs have different minimum central angles to the serving orbit, resulting in different minimum possible horizontal distances to the orbit. Consequently, $\XMIN$ is a random variable. However, due to its small variance, $\XMIN$ can be approximated with a constant value, $\hat{X}_{\mathrm{min}}$, which represents the minimum horizontal distance from point $C$ to the nearest point on the serving orbit of a given cell. This approximation, which is validated in~\cite[Theorem 2]{doppler_tanash}, provides an effective tool to model the Doppler shift distribution.


\vspace{-.3cm}
\subsection{Channel and Signal Models}

In this letter, we assume that the satellite's elevated position ensures a dominant line-of-sight (LoS) link to the user, such that multipath components are neglected. Consequently, no multipath-induced Doppler spread is considered. After common compensation, the remaining impairment is a geometry-dependent residual Doppler frequency offset which, due to the short OFDM symbol duration, is assumed constant within each symbol. 
According to 3GPP TR 38.821~\cite{TR38.821}, the minimum elevation angle is set at $\alphamin = 10^\circ$, below which communication is considered impossible. This channel model is ideally suited for suburban and rural scenarios, where the LoS probability is $78.2\%$ for $\alphamin = 10^\circ$~\cite[Table 6.6.1-1]{TR38.821}.


{This study leverages the OFDM model introduced in~\cite{sinc_long} to characterize both the useful signal power and the inter-carrier interference (ICI) arising from the loss of subcarrier orthogonality, which in the present system is induced by residual Doppler shift. Although the model in~\cite{sinc_long} was originally developed to describe ICI caused by Doppler spread due to rapid channel variation, the fundamental origin of orthogonality loss in OFDM is a frequency mismatch over the symbol duration, regardless of its physical cause.
In the considered LEO OFDMA system, Doppler compensation is performed at a single reference point, resulting in user-dependent residual Doppler shifts across the satellite footprint. During the reception of an OFDM symbol, these residual Doppler shifts appear as user-specific frequency mismatches with respect to the OFDM subcarrier grid, reflecting the underlying satellite geometry. While the physical origin of this impairment differs from classical Doppler spread, its impact on subcarrier orthogonality, and hence on useful signal attenuation and ICI generation, is identical. This observation allows the residual Doppler impairment to be embedded into the OFDM ICI framework of~\cite{sinc_long} and, crucially, propagated into the SINR and coverage probability analysis.}
In particular, the useful signal power $\Omega$ and ICI $I$ can be calculated for an infinite number of subcarriers according to \cite[Eq. 5]{sinc_long} as
\begin{align}
\label{eq:useful_signal_kaiser}
\Omega&=\abs{\eta}^2\sinc^2(\fd T)\\
\label{eq:ICI_kaiser}
I&= \abs{\eta}^2 \big(1-\sinc^2(\fd T)\big),
\end{align}
with the link attenuation $\eta$ defined as
\begin{align}
\label{eq:path_loss}
\eta={\sqrt{l\,\GR\GMAX}\frac{\lambda}{4\pi D_t}}={\sqrt{l\,\GR\GMAX}\frac{\lambda}{4\pi\sqrt{X_t^2+\HS^2}}}.
\end{align}
Above, $\fd$ is the Doppler shift magnitude defined in (\ref{eq:doppler_s_approx}), $T$ is the OFDM symbol duration, calculated as $T=\frac{1}{\Delta f}$, with $\Delta f$ representing the subcarrier spacing. Additionally, $l$, $\lambda$, $\GR$, and $\GMAX$ denote, respectively, the rain attenuation gain, the carrier wavelength, the ground terminal antenna gain, and the maximum satellite antenna gain.
The ICI in (\ref{eq:ICI_kaiser}) is accurate for subcarrier counts above $60$, ideal for OFDM satellite systems that typically operate with $256$ or more subcarriers.
It is important to note that although (\ref{eq:useful_signal_kaiser}) and (\ref{eq:ICI_kaiser}) are derived for the central subcarrier, they still hold for the rest of the subcarriers since those near the edge of the OFDM symbol suffer from less ICI. Therefore, (\ref{eq:useful_signal_kaiser}) and (\ref{eq:ICI_kaiser}) represent the worst-case scenario with the maximum possible ICI. 

In the presence of Doppler shift, the instantaneous SINR at any ground terminal within $S(C,R_c)$ at time $t$, can be calculated using (\ref{eq:useful_signal_kaiser}) and (\ref{eq:ICI_kaiser}) as 
\begin{align}
\label{eq:SINR_short}
\vspace{-.7cm}
&\mathrm{SINR}=\frac{\PTX\,\Omega}{\PTX\,I+N_o},
\end{align}
where $\PTX$ denotes the transmitted power of the satellite.
The transmitter precompensates the Doppler shift at a reference point in a cell, and
the residual Doppler shift represents the effective component of the Doppler shift that impacts the SINR and is denoted herein as $\Delta \delta_t$. In particular, the Doppler shift $\fd$ can be written as $\fd= \delta_t+\Delta \delta_t$, for which $ \delta_t=\fd(\hat{X}_t,\hat{X}_{\mathrm{min}})$ is the common part of Doppler shift experienced by all the ground terminals in the cell, regarded as a constant at a given time instant, and is calculated using (\ref{eq:doppler_s_approx}) at the reference point, which is the cell's center herein. Therefore, in the presence of the residual Doppler shift $\Delta \delta_t(X_t,\XMIN)=\fd(X_t,\XMIN)-\fd(\hat{X}_t,\hat{X}_{\mathrm{min}})$, the instantaneous SINR at any ground terminal within $S(C,R_c)$ at time $t$, can be calculated as
\begin{equation}
\resizebox{1.0\hsize}{!}{$\begin{aligned}\label{eq:SINR}
&\mathrm{SINR}=\frac{{A }\sinc^2(\Delta \delta_t(X_t,\XMIN) T)\, ({X_t^2+\HS^2})^{-1}}{{A} \big(1-\sinc^2(\Delta \delta_t(X_t,\XMIN) T)\big)\,({X_t^2+\HS^2})^{-1}+N_o},
\end{aligned}$}
\end{equation}
where $A=\PTX\,l\,\GR\,\GMAX\,\left(\frac{\lambda}{4\,\pi}\right)^2$.  




\section{PERFORMANCE ANALYSIS}
\label{sec:PERFORMANCE ANALYSIS}
This section employs the SINR derived in Section~\ref{sec:system_model} to calculate the downlink coverage probability for an arbitrarily located ground terminal within a cell $S(C,R_c)$ on the Earth's surface. It accounts for the residual Doppler effect among the terminals in the cell, which causes ICI.

\begin{figure*}[h]
\vspace{-.5cm}
	\begin{center}
	  		\subfigure[Impact of frequency band of operation on coverage probability ]{\includegraphics[trim=.1cm .1cm .1cm .1cm, clip=true,width=0.47\textwidth]{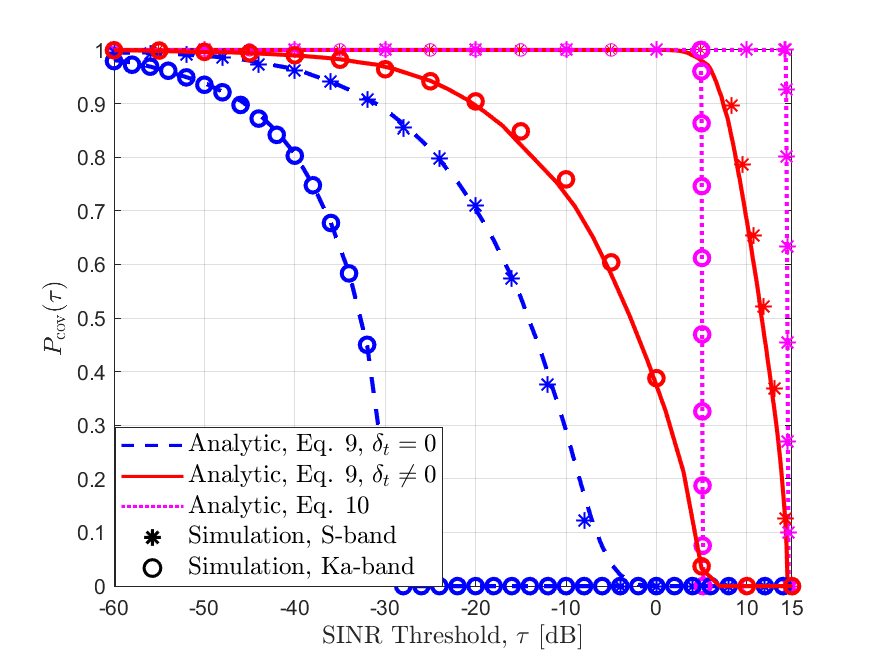}}
		\hfill     \vspace{.65cm}
		\subfigure[Impact of subcarrier spacing on coverage probability in S-band]{\includegraphics[trim=.1cm .1cm .1cm .1cm, clip=true,width=0.49\textwidth]{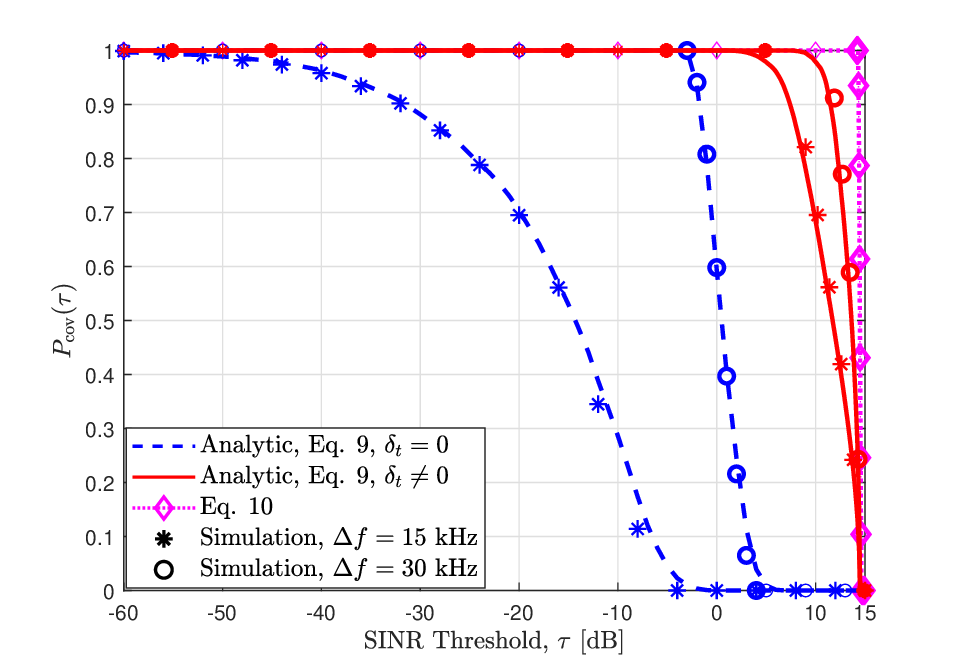}}\\[-1cm]
		\subfigure[Impact of cell size on coverage probability]{\includegraphics[trim=.1cm .04cm .1cm .1cm, clip=true,width=0.49\textwidth]{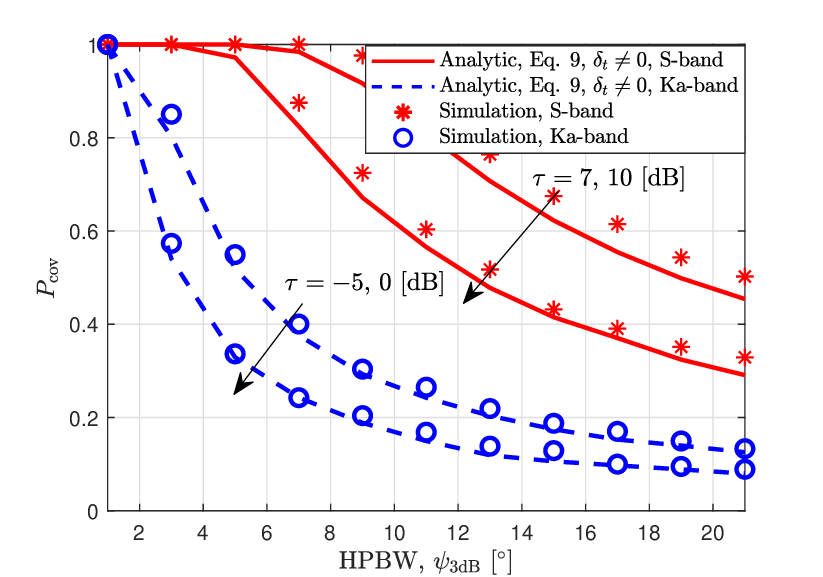}}
		\hfill      \vspace{.8cm}
		\subfigure[Impact of satellite location on coverage probability in Ka-band ]{\includegraphics[width=0.49\textwidth]{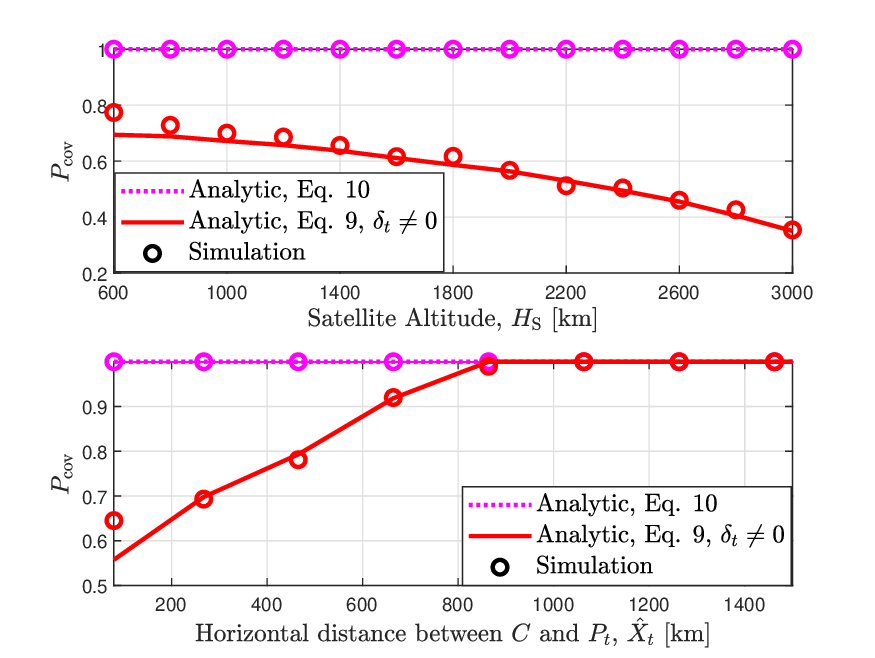}}
      \vspace{-.7cm}
\caption{Coverage probability under varying parameters, including frequency band, spacing, cell size, and satellite location. Eq.~\ref{eq:cov_prob} with $\delta_t = 0$ indicates no Doppler compensation; Eq.~\ref{eq:cov_prob} with $\delta_t \ne 0$ represents residual Doppler after canceling the common component; and Eq.~\ref{eq:ideal_cov} corresponds to the ideal case with fully compensated Doppler.}
        \label{fig:only_fig}
    \end{center}
    \vspace{-.8cm}
\end{figure*}

The coverage probability for an arbitrarily located ground terminal within $S(C,R_c)$ that is served by a LEO satellite in a network affected by the residual Doppler shift is given by (\ref{eq:cov_prob})\footnote{{Small-scale fading or shadowing can be incorporated by conditioning the SINR expression in (9) on the instantaneous channel gain and averaging the resulting coverage probability over the corresponding channel distribution, assuming the channel remains constant over the OFDM symbol duration. This extension preserves the structure of the analysis and can be evaluated numerically.}}, shown at the bottom of the previous page in which (a) follows from (\ref{eq:doppler_s_approx}) by approximating $\XMIN$ with the constant value $\hat{X}_{\mathrm{min}}$, while $F_{X_t}(\cdot)$ is given in (\ref{eq:CDF_Z}). It is important to note that while (\ref{eq:cov_prob}) accounts for the residual Doppler shift after compensating the common component ($\delta_t \ne 0$), it also remains applicable for evaluating the coverage probability without any Doppler compensation by setting $\delta_t = 0$ in (\ref{eq:cov_prob}), under which the residual Doppler shift equals the Doppler shift magnitude at the ground terminal, i.e., $\Delta \delta_t(X_t, \XMIN) = \fd(X_t, \XMIN)$.

In networks with fully compensated Doppler shift where zero residual frequency offset occurs at the ground terminal, whether through ideal compensation algorithms or in scenarios where Doppler effects are inherently negligible, the coverage probability is given by \stepcounter{equation}
\begin{align} 
\label{eq:ideal_cov}
P_{\mathrm{cov}}(\tau)
&=\mathbb{P}\bigg(\frac{{A }}{({X_t^2+\HS^2})\,N_o}>\tau\bigg)\nonumber\\
&=F_{X_t}\Big(\sqrt{\frac{A}{\tau N_o}-\HS^2}, R_c \mid \hat{X}_t\Big).
\end{align}
The coverage probability in (\ref{eq:ideal_cov}) serves as an optimal theoretical bound, benchmarking systems with uncompensated or partially compensated Doppler shifts and highlighting the performance gap between ideal and practical mobile scenarios with residual Doppler effects.

\section{NUMERICAL RESULTS}
\label{sec:Numerical Results}
In this section, we validate the coverage probability expressions derived in Section~\ref{sec:PERFORMANCE ANALYSIS} through Monte--Carlo simulations. We also study the effect of different system parameters on its performance. For our simulations, we utilize two sets of satellite parameters that adhere to the 3GPP standards detailed in \cite[Tables 6.1.1.1-1 and 6.1.1.1-2]{TR38.821} along with two types of ground terminal characteristics described in \cite[Table 6.1.1.1-3]{TR38.821}. These parameters serve as the baseline for our analysis. Specifically, the first set considers communication over the S-band with a carrier frequency $f_o = 2$ GHz, a satellite equivalent isotropic radiated power (EIRP) density of $28$ dBW/MHz, bandwidth$=30$ MHz, a maximum satellite gain of $\GMAX=24$ dBi with a half-power beamwidth $\psi_{3\mathrm{dB}} = 8.832^\circ$. The ground terminals are handheld devices equipped with omnidirectional antennas. The second set considers communication over the Ka-band with $f_o = 20$ GHz, a satellite EIRP density of $-4$ dBW/MHz, bandwidth$=400$ MHz, $\GMAX=30.5$ dBi, $\psi_{3\mathrm{dB}} = 4.4127^\circ$, and a very small aperture ground terminal (VSAT) of gain $\GR=39.7$ dBi. For both parameter sets, referred to herein as S-band and Ka-band scenarios respectively, we assume $i=53^\circ$, $\alphamax=85^\circ$, $\HS=600$ km and $l=-3.125$ dB, unless otherwise stated.

Fig. \ref{fig:only_fig}(a) shows the coverage probability derived in (\ref{eq:cov_prob}) in the presence of residual Doppler shift, which persists after compensating for the common Doppler component at a reference point. The results are compared to scenarios without frequency compensation of the common component and the ideal case where Doppler shift is either fully compensated or neglected, as given in (\ref{eq:ideal_cov}). The simulations confirm the high accuracy of the derived coverage probability, where they closely match with the analytical expressions in (\ref{eq:cov_prob}) and (\ref{eq:ideal_cov}) across different threshold values for both the S-band and Ka-band, validating the derived models.
The figure highlights the substantial impact of Doppler shift on performance, even after partial compensation, demonstrating that residual Doppler still degrades coverage compared to the ideal case. This emphasizes the need for precise Doppler modeling, especially in higher-frequency bands (e.g., Ka-band), where Doppler variations are more severe.
Furthermore, Fig.~\ref{fig:only_fig}(b), demonstrates that increasing the subcarrier spacing in OFDM symbols reduces the effect of Doppler shift and makes the system more robust to frequency variations, thereby decreasing intercarrier interference and improving system performance. The impact of the cell's size containing the ground terminals is illustrated in Fig.~\ref{fig:only_fig}(c). As the HPBW increases, and consequently, the cell radius expands, the Doppler shift variation across the ground terminals becomes more pronounced, leading to a reduction in coverage probability for both S-band and Ka-band scenarios.

In Fig.~\ref{fig:only_fig}(d), the impact of the location of the satellite is depicted in terms of the satellite altitude and the horizontal distance $\hat{X}_t$ between the cell's center and the satellite’s subpoint $P_t$. In the presence of residual Doppler shift, increasing the satellite altitude reduces its velocity component along the LoS path to the ground terminal, thereby decreasing the overall Doppler shift. Nevertheless, as the altitude increases, the resulting longer slant distance leads to greater path loss, which becomes the dominant factor causing a decline in coverage probability.
Conversely, at a fixed satellite altitude (e.g., 
$\HS=1200$ km herein), increasing the horizontal separation between the cell and the serving LEO satellite improves the coverage probability. This improvement is primarily attributed to the reduction in residual Doppler shift as the elevation angle decreases, a behavior also confirmed in~\cite[Fig. 10] {doppler_tanash}. The reduced residual Doppler distortion reduces ICI, 
thus improving coverage performance, despite the relatively increased path loss associated with longer slant distances.

\vspace{-.3cm}
\section{CONCLUSION}
This paper provided a comprehensive analysis of residual Doppler shift's impact on LEO satellite-terrestrial systems, particularly focusing on coverage probability. By accurately modeling the Doppler shift, characterizing critical distances, and deriving the ICI, we offered critical insights into how various system parameters influence OFDM performance under realistic Doppler conditions.
The results demonstrated that residual Doppler shift substantially degrades coverage, even after compensating for the common Doppler component, with more pronounced effects at higher frequency bands. Both analytical and simulation results showed that design factors such as OFDM subcarrier spacing, antenna beamwidth, satellite altitude, and user-satellite geometry can critically influence the induced Doppler shift and its impact on system performance.






\vspace{-.3cm}

\bibliographystyle{IEEEtran}
\bibliography{refs}

\end{document}